\documentclass{webofc}
\usepackage[varg]{txfonts}   
\begin{document}
\title{Asymptotic symmetries and soft theorems in higher-dimensional gravity}
%
%

\author{\firstname{Stefano} \lastname{Lionetti}\inst{1}\fnsep\thanks{\email{stefano.lionetti@unisalento.it}}        
}

\institute{Dipartimento di Matematica e Fisica "Ennio De Giorgi", Università del Salento and INFN-Lecce,
Via Arnesano, 73100 Lecce, Italy
          }

\abstract{%
Soft theorems can be recast as Ward identities of asymptotic symmetries.
We review such relation for the leading and subleading soft graviton theorems in arbitrary even dimensions. While soft theorems are trivially generalized to dimensions higher than four,
the charges of asymptotic symmetries are plagued by divergences requiring a renormalization. 
We argue that the renormalized charges of these symmetries can be determined by rewriting soft theorems as Ward identities. 
In order to show that the charges of such identities generate asymptotic symmetries, we propose a suitable commutation relation among certain components of
the metric fields
}
\maketitle

\section{Introduction}
\label{intro}

Over the last few years a triangular equivalence relation was discovered connecting three apparently different topics: asymptotic symmetries, soft theorems and memory effects \cite{Strominger:2013jfa,He:2014laa,Strominger:2014pwa}. This equivalence relation can be drawn potentially in every theory with a massless particle, for example in QED, QCD, SUSY and gravity (for a review of the topic, see \cite{Strominger:2017zoo}). 
In this paper we focus on the relationship between asymptotic symmetries and soft theorems at leading and subleading order in gravity in arbitrary even dimensions. 

In gravity the asymptotic symmetries are diffeomorphisms that preserve the asymptotic flatness of spacetime \cite{Bondi:1962px,Sachs:1962wk,Sachs:1962zza}. Such transformations map an asymptotically flat metric into an other asymptotically flat metric. 
The asymptotic symmetry group is an infinite-dimensional extension of the Poincaré group.
The Ward identitity of asymptotic symmetries turn out to be nothing but soft theorems. 

The soft graviton theorem \cite{Weinberg:1965nx,Cachazo:2014fwa} is a universal formula relating scattering amplitudes that differ only by the addition of a graviton whose energy $\omega$ is taken to zero
\begin{equation}
\mathcal{M}_{n+n'+1}\left(q=\omega \hat{q} ; p_{1}, \ldots, p_{n+n'}\right) = \left[S^{(1)}+S^{(2)}\right] \mathcal{M}_{n+n'}\left(p_{1}, \ldots, p_{n+n'}\right)+\mathcal{O}\left(\omega \right)
\end{equation}
where $\{p_1, \ldots p_n\}$ are the momenta of the incoming particles, while $\{p_{n+1}, \ldots p_{n+n'}\}$ are the momenta of the outgoing particles. $S^{(1)}$ and $S^{(2)}$ are given by 
\begin{equation}
\begin{aligned} S^{(1)} & \equiv \frac{\kappa}{2} \left( \sum_{k=n+1}^{n+n'} \frac{\varepsilon_{\mu \nu} p_{k}^{\mu} p_{k}^{\nu}}{p_{k} \cdot q} - \sum_{k=1}^{n} \frac{\varepsilon_{\mu \nu} p_{k}^{\mu} p_{k}^{\nu}}{p_{k} \cdot q}\right)\\  
S^{(2)} & \equiv-\frac{i \kappa}{2} \left(  \sum_{k=n+1}^{n+n'} \frac{\varepsilon_{\mu \nu} p_{k}^{\mu} q_{\rho} J_{k}^{ \rho \nu }}{p_{k} \cdot q}- \sum_{k=1}^{n} \frac{\varepsilon_{\mu \nu} p_{k}^{\mu} q_{\rho} J_{k}^{ \rho \nu }}{p_{k} \cdot q} \right) \end{aligned}
\end{equation}
where $\kappa^2=32 \pi G$ and $J_{k}^{ \rho \nu }$  is the total angular momentum (orbital+spin) of the $k$-th particle and $\varepsilon_{\mu \nu}$ is the polarization tensor of the graviton.
In the soft limit $\omega \rightarrow 0$, the leading term given by
  $S^{(1)}$ is of order $1/\omega$, while the subleading term given by
  $S^{(2)}$ is constant in $\omega$.
The formula indeed comprises the leading and subleading soft graviton
theorem. We are not concerned with higher order terms in our analysis.

In four dimensions the leading soft theorem is equivalent to the supertranslation Ward identity \cite{Strominger:2013jfa,He:2014laa}, while the subleading soft theorem is equivalent to the Diff$(S^2)$ Ward identity \cite{Campiglia:2014yka,Campiglia:2015yka}. When computing the Diff$(S^2)$ charges, we encounter divergences and we need to take care of them by adding appropriate counterterms to the action as shown in \cite{Compere:2018ylh,Compere:2020lrt,Chandrasekaran:2021vyu}. As we will see, in $d>4$ such divergences are even more present (supertranslations charges will require a renormalization too). Together, supertranslations and Diff$(S^2)$ make up the asymptotic symmetry group as defined in \cite{Campiglia:2014yka,Campiglia:2015yka}
\begin{equation}
\mathcal{AG}=\operatorname{Diff}\left(S^2\right) \ltimes \text { Supertranslations }
\end{equation}

The equivalence relation between asymptotic symmetries and soft theorems seems to be somewhat more obscure in dimensions higher than four. Indeed soft gravitons theorems hold in any dimension $d=2m+2$ while both supertranslations and Diff$(S^{2m})$ charges seem to diverge in $d>4$. Many researchers avoided such divergences' problem by imposing strong falloff conditions, leaving only the Poincar\'e group as part of the asymptotic symmetry group \cite{Hollands:2016oma,Hollands:2003ie,Tanabe:2011es}. 
In $d=4$, such argument would not hold. Falloff conditions disallowing supertranslations and Diff$(S^2)$ automatically exclude all the generic radiative solutions in $d=4$.
In any case, we will not consider such strong falloff conditions in $d>4$ either, since our aim is to preserve the equivalence relation between asymptotic symmetries and soft theorems.
However, a renormalization seems to be mandatory in order to solve the divergence problems.\footnote{In \cite{Aggarwal:2018ilg} the author proposed boundary conditions that eliminate divergences in the supertranslations charges. However these boundary conditions are only consistent in linearized gravity and a renormalization is still needed in the full theory. }

In this paper we present the proper asymptotic symmetry group that can be linked to soft graviton theorems. We then rewrite the leading and subleading soft graviton theorems as Ward identities.
We argue
that such identities are associated to asymptotic symmetries by proposing a suitable commutation relation among certain components of the metric fields. We expect the need for a renormalization in order to prove such commutation relation.
We work in the Bondi gauge and linearized gravity coupled to massless matter throughout the paper. Our discussion is restricted to tree-level.
Asymptotic symmetries have also been studied in higher dimensions in the full nonlinear theory, for any spin, in odd-dimensional cases and in asymptotically (A)dS spacetimes \cite{Campoleoni:2020ejn,Capone:2021ouo,Fuentealba:2022yqt,Bekaert:2022ipg,Fiorucci:2020xto,Chowdhury:2022nus}.

\section{Asymptotic symmetries in higher dimensions}
In gravity the asymptotic symmetries are diffeomorphisms that preserve the asymptotic flatness of spacetime.
In order to define precisely asymptotic flatness, we need to specify the rate at which the metric approaches a Minkowski metric at asymptotically large distances. Unfortunately there is no unambiguous method of determining such asymptotic falloff conditions which are often only a posteriori justified. There are however guidelines that must be followed:
the falloff conditions should be weak enough so that all interesting solutions are allowed, but strong enough to rule out unphysical solutions, such as those with infinite energy.
Various options are discussed in the literature leading to different results \cite{Hollands:2016oma,Pate:2017fgt,Kapec:2015vwa,Colferai:2020rte}. In this paper, we advocate for falloff conditions leading to non-trivial relations among $\mathcal{S}-$matrix elements.
Indeed Ward identities, associated to proper asymptotic symmetries, turn out to be nothing but soft theorems. 

We are interested in asymptotically flat spacetimes at both future and past null infinity $\mathcal{I}^{\pm}$.
For concreteness, let us focus on future null infinity.
We choose the coordinate system $(u,r,z^a)$, where $u=t-r$ is the retarded time, $r$ is the radial coordinate and $z^a$ ${(a=1,\dots,2m)}$ are the coordinates on the sphere $S^{2m}$. We work in the Bondi gauge imposing the following $2m+2$ conditions
\begin{equation} \label{eq:gaugefixcond}
g_{r r}=0, \quad \quad g_{r a}=0, \quad \quad
\operatorname{det} g_{a b}=r^{4 m} \operatorname{det} \gamma_{a b}
\end{equation}
where $\gamma_{ab}$ is the unit $S^{2m}$ sphere metric\footnote{We are not assuming that $\gamma_{ab}$ is the standard round sphere metric.} with covariant derivative $D_a$. All angular indices $(a,b,c\dots)$ are raised and lowered with $\gamma_{ab}$. 

We can now define an asymptotically flat spacetime by imposing the following falloff conditions on the metric
\begin{equation} \label{eq:asympticcondition}
g_{uu}=\mathcal{O}(1), \quad \quad g_{u r}=-1+\mathcal{O}\left(\frac{1}{r}\right), \quad \quad g_{u a}=\mathcal{O}(1), \quad \quad
g_{a b}=\mathcal{O}(r^2) 
\end{equation}
The asymptotic symmetry group is the group of all non-trivial diffeomorphisms preserving the gauge-fixing conditions \eqref{eq:gaugefixcond} and the asymptotic falloffs \eqref{eq:asympticcondition}. Such diffeomorphisms are generated by the vector
\begin{equation} \label{eq:vectbmsdimgentraslrot}
  \begin{aligned}
    \zeta^{u}&=f(z)+\frac{u}{2 m} D_{a} Y^{a}(z)+\ldots \\
    \zeta^{a}&=-\frac{1}{r}D^af(z)+Y^{a}(z)-\frac{u}{2mr}D^aD_bY^b(z)+\ldots \\
    \zeta^{r}&=\frac{1}{2 m}D^2f(z)-\frac{r}{2 m} D_{a} Y^{a}(z)+\frac{u}{4 m^2}D^2 D_{a} Y^{a}(z)+\ldots
  \end{aligned}
\end{equation}
Transformations with $Y^{a}=0$ and an arbitrary function $f$ on the sphere are known as supertranslations and they are a generalization of the global translations.
Transformations with $f=0$ and an arbitrary smooth vector $Y^{a}$ on the sphere are known as Diff$(S^{2m})$ and they are a generalization of the global Lorentz transformations. 
Note that not all $\zeta$ are asymptotic killing vector fields
\begin{equation}
\lim_{r \rightarrow \infty} \nabla_{(a} \zeta_{b)} \neq 0
\end{equation}
but they all are asymptotically divergence-free 
\begin{equation}
\lim_{r \rightarrow \infty} \nabla_a \zeta^a = 0
\end{equation}

In $d=4$ an alternative and more restrictive description of asymptotic flatness has been previously proposed, leading to the requirement that $Y$ is a conformal killing vector (CKV) of the sphere \cite{Barnich:2010ojg,Barnich:2011mi,Kapec:2014opa}.
In $d=4$, there are infinitely many local solutions $Y$ to the CKV equation, corresponding to infinitely many transformations generated by $\zeta$. Such transformations are called superrotations and they generalize the Lorentz transformations given by global solutions to the CKV equation.
However the number of independent constraints, imposed by the CKV equation, grows with the dimensions of the space. In $d=4$ the CKV equations are the Cauchy-Riemann equations and so there are infinite independent local solutions. For $d>4$, the system of equations is over-determined and there are only a finite number of independent solutions: the global Lorentz transformations.
In dimensions higher than four, since one cannot extend Lorentz transformations to superrotations, the Diff$(S^{2m})$ seem to be the proper asymptotic symmetries to link to the subleading soft graviton theorem. Therefore, here we will consider arbitrary smooth vectors $Y_a$ of the sphere.
The relationship between superrotations and Diff$(S^{2})$ in $d=4$ has been studied in \cite{Donnay:2020guq} and it involves a shadow transform. For $d>4$ such shadow transform is not well defined and this is consistent with the fact that superrotations don't exist in higher dimensions.

\section{From soft theorems to asymptotic symmetries}
Let us start by considering the leading soft graviton theorem
\begin{equation} \label{eq:lsgt}
\lim _{\omega \rightarrow 0} \mathcal{M}_{n+n'+1}\left(q=\omega \hat{q} ; p_{1}, \ldots, p_{n+n'}\right) = S^{(1)} \mathcal{M}_{n+n'}\left(p_{1}, \ldots, p_{n+n'}\right)
\end{equation}
where
\begin{equation}
S^{(1)}  \equiv \frac{\kappa}{2} \left( \sum_{k=n+1}^{n+n'} \frac{\varepsilon_{\mu \nu} p_{k}^{\mu} p_{k}^{\nu}}{p_{k} \cdot q} - \sum_{k=1}^{n} \frac{\varepsilon_{\mu \nu} p_{k}^{\mu} p_{k}^{\nu}}{p_{k} \cdot q}\right)\\  
\end{equation}
Such theorem can be recast into a Ward identity (for the derivation see \cite{Kapec:2015vwa})
\begin{equation}
\langle out |Q^+\mathcal{S}-\mathcal{S}Q^-| in \rangle  =0
\end{equation}
where the charges can be decomposed into a soft and a hard part
\begin{equation}
Q^{\pm}=Q_{H}^{\pm}+Q_{S}^{\pm}
\end{equation}
The hard part comes from the soft factor $S^{(1)} $ on right-hand side of eq \eqref{eq:lsgt} after some manipulation, while the soft part comes from the left-hand side of the equation. Therefore, $Q_H$ generates asymptotic symmetries on the external hard states, while $Q_S$ is responsible for the creation of a soft graviton. 
Focusing for now on $Q^+$, the hard charge takes the form
\begin{equation}
\left\langle z_{n+1}, \ldots\right| Q_{H}^{+}=\left\langle z_{n+1}, \ldots\right| \sum_{k=n+1}^{n+n^{\prime}} E_{k} f\left(z_{k}\right)
\end{equation}
which represent the action of supertranslations on each outgoing particle. The soft charge is given by 
\begin{equation}
Q_{S}^{+}
\propto 
\int d^{2 m} z \sqrt{\gamma} f(z) \prod_{l=m+1}^{2 m-1}\left(D^{2}-(2 m-l)(l-1)\right)I^{(m-2)} D^{a} D^{b} g_{a b}^{(m-2)}
\end{equation}
where $g^{(n)}_{\mu \nu}$ is the $n$-component of the metric in the $1/r$ expansion and we defined the integral operator $I^{(m-2)}=\left(\int d u\right)^{(m-2)}$. \footnote{In the four-dimensional case $(m=1)$, we use the following prescriptions: $\prod_{l=m+1}^{2m-1} \left(D^{2}-(l-1)(2 m-l) \right) = 1$ and $I^{(m-2)}=\partial_u$.} 
In \cite{Kapec:2015vwa,Colferai:2020rte} the following commutation relation has been proposed
\begin{equation} \label{eq:mycomrel}
\left[\left.g_{uu}^{(2 m-1)}(z)\right|_{\mathcal{I}_{-}^{+}}, C\left(u^{\prime}, z^{\prime}\right)\right]=\frac{8 \pi i G}{m} \frac{\delta^{2 m}\left(z-z^{\prime}\right)}{\sqrt{\gamma}}
\end{equation}
where 
$
g_{a b}^{(-1)}=\left(\frac{1}{m} \gamma_{a b} D^2 -2D_a D_b\right) C 
$. Thanks to this commutation relation, it is possible to show that the charges $Q=Q_S+Q_H$ generate supertanslations. The eq. \eqref{eq:mycomrel} has not been proven yet but it is consistent with the four-dimensional case \cite{He:2014laa}.
In order to derive the formula in higher dimensions, a renormalization will be required. 

Let us now consider the subleading soft graviton theorem 
\begin{equation}
\lim _{\omega \rightarrow 0}\left(1+\omega \partial_{\omega}\right)\mathcal{M}_{n+n'+1}\left(q=\omega \hat{q} ; p_{1}, \ldots, p_{n+n'}\right) = S^{(2)} \mathcal{M}_{n+n'}\left(p_{1}, \ldots, p_{n+n'}\right)\end{equation}
where 
\begin{equation}
S^{(2)}=-\frac{i\kappa}{2}\left[ \sum_{k=n+1}^{n+n'} \frac{p_{k \mu} \varepsilon^{\mu \nu} q^{\lambda} J_{k \lambda \nu}}{p_{k} \cdot q}-\sum_{k=1}^{n} \frac{p_{k \mu} \varepsilon^{\mu \nu} q^{\lambda} J_{k \lambda \nu}}{p_{k} \cdot q}\right]
\end{equation}
The operator $\left(1+\omega \partial_{\omega}\right)$ projects out the pole $\frac{1}{\omega}$ at the leading order in $ S^{(1)}$. 

The subleading soft theorem can also be recast as a Ward identity (for the derivation see \cite{Colferai:2020rte}). The hard charge of such identity is given by 
\begin{equation} \label{eq:qharddateosofficeaaaa}
\begin{aligned} \left\langle out \right|  Q_{H}^{+} =    i \sum_{k=n+1}^{n+n'} \left(Y^{a}\left(z_{k}\right) \partial_{z^a_{k}}-\frac{E_{k}}{2m} D_{a} Y^{a}\left(z_{k}\right) \partial_{E_{k}}\right)\left\langle out \right| \end{aligned}
\end{equation}
which represent the action of a Diff$(S^{2m})$ on each outgoing particle.
The soft charge is given by
\begin{equation}
\begin{aligned}
Q^+_S \propto& \int du \int d^{2m}z  \sqrt{\gamma}\text{ }u D.Y  \prod_{l=m+1}^{2m-1} \left(D^{2}-(l-1)(2 m-l) \right)  I^{(m-2)}\left( D^aD^bg_{a b}^{(m-2)} \right)
\end{aligned}
\end{equation}
Using again the commutation relation \eqref{eq:mycomrel}, we are able to show that such charges generate Diff$(S^{2m})$.

\section{Conclusions and outlook} 
We examined the relation between asymptotic symmetries and soft theorems at leading and subleading order in higher-dimensional gravity. 
In order to preserve the equivalence relation in $d>4$, the asymptotic charges need to be renormalized. Such renormalization procedure has already been proven to be effective in $d=4$ in \cite{Compere:2018ylh,Compere:2020lrt,Chandrasekaran:2021vyu}. The search for the right counterterms in $d>4$ is currently an open problem that needs further investigation. However we already know the final form of the renormalized charges since we can derive them from soft theorems. 
Indeed, we can recast the soft graviton theorems as Ward Identities.
We then argued that such identities are associated to asymptotic symmetries, provided a suitable commutation relation between metric fields holds. 
Recent analyses also support the form of asymptotic charges we derived \cite{Campoleoni:2020ejn}.
As a result, supertranslations and Diff($S^{2m}$) are symmetries of gravitational scattering.

There are still many other open topics that need to be tackled in the future. Soft theorems and asymptotic symmetries are part of a triangular equivalence relation called IR Triangle. The last corner of the triangle is a memory effect and it represents an exciting experimental prospect for the coming future of the gravitational wave physics. The spin memory effect has been studied in relation to the subleading soft graviton in $d=4$ in \cite{Pasterski:2015tva}.  It would be
interesting to examine such effect in higher dimensions.

Moreover asymptotic symmetries and soft theorems play a central role in the celestial holography program. 
Celestial holography proposes a duality between gravitational scattering in asymptotically flat spacetimes and a conformal field theory living in two dimensions lower on the celestial sphere (for reviews of the topic see \cite{Pasterski:2021rjz, Raclariu:2021zjz,McLoughlin:2022ljp}). This field of research has been extensively studied in $d=4$ but it can also be generalized to higher dimensions \cite{Kapec:2017gsg,Pasterski:2017kqt} where further investigations are needed.

Lastly, the sub-subleading soft graviton theorem has been recently studied in relation to
asymptotic symmetries in $d=4$ \cite{Campiglia:2016efb,Campiglia:2016jdj,Freidel:2021dfs}. It would be
worthwhile to also extend such analyses to arbitrary dimensions.
\section*{Acknowledgements}
We thank C. Corianò for collaborating to related analysis.
The work of S.L. is partly supported by INFN Iniziativa Specifica QFT-HEP.

%
%
%

\end{document}